\documentclass[twocolumn,journal]{IEEEtran}
\usepackage{amssymb,color}
\usepackage{amsmath}
\usepackage{bbm}
\usepackage{graphicx}
\usepackage{epstopdf}
\usepackage{amsthm,amsfonts}
\usepackage{subcaption}
\usepackage{accents}

\usepackage{geometry}
\usepackage{amssymb,color,multirow,xcolor,colortbl}
\usepackage{amsmath}
\usepackage{tcolorbox}
\usepackage{scalefnt}
\usepackage{comment}

\usepackage{algorithm}
\usepackage{algorithmic}
\usepackage{psfrag}
\usepackage{cite}
\usepackage{pifont}
\usepackage[font=footnotesize, skip=0pt]{caption}
\allowdisplaybreaks

\makeatletter
\newcommand{\oset}[3][0ex]{%
  \mathrel{\mathop{#3}\limits^{
    \vbox to#1{\kern-2\ex@
    \hbox{$\scriptstyle#2$}\vss}}}}
\makeatother

\usepackage[acronym,nonumberlist,style=super,toc]{glossaries}

\makenoidxglossaries
\newacronym{5g}{5G}{fifth generation}

\newacronym{awgn}{AWGN}{additive white Gaussian noise}
\newacronym{amc}{AMC}{adaptive modulation and coding}

\newacronym{ber}{BER}{bit error rate}
\newacronym{bler}{BLER}{block error rate}
\newacronym{bpsk}{BPSK}{binary phase-shift keying}
\newacronym{bs}{BS}{base station}
\newacronym{bch}{BCH}{Bose–Chaudhuri–Hocquenghem}

\newacronym{csi}{CSI}{channel state information}

\newacronym{dl}{DL}{downlink}

\newacronym{et}{throughput}{throughput}
\newacronym{apa}{APA}{adaptive power assignment}
\newacronym{fpa}{FPA}{fixed power assignment}
\newacronym{apt}{aTp}{average throughput}
\newacronym{ipt}{iTp}{instantaneous throughput}

\newacronym{iid}{iid}{independent and identically distributed}
\newacronym{iot}{IoT}{Internet of Things}
\newacronym{irs}{IRS}{intelligent reflecting surface}

\newacronym{jmld}{JMLD}{joint-multiuser maximum likelihood detector}

\newacronym{lut}{LUT}{look-up table}

\newacronym{mpsk}{$M$-PSK}{M-ary phase shift keying}
\newacronym{mimo}{MIMO}{multiple-input-multiple-output}

\newacronym{noma}{NOMA}{non-orthogonal multiple access}

\newacronym{oma}{OMA}{orthogonal multiple access}

\newacronym{pdf}{PDF}{probability density function}
\newacronym{pa}{PA}{power assignment}
\newacronym{pcb}{PCB}{power coefficient bound}

\newacronym{qos}{QoS}{quality of service}
\newacronym{qpsk}{QPSK}{quadrature phase-shift keying}
\newacronym{qam}{$M$-QAM}{$M$-ary quadrature amplitude modulation}
\newacronym{mqam}{QAM}{quadrature amplitude modulation}

\newacronym{rsr}{RSR}{reliable sum rate}
\newacronym{ser}{SER}{symbol error rate}

\newacronym{se}{SE}{spectral efficiency}
\newacronym{sc}{SC}{superposition coding}
\newacronym{sic}{SIC}{successive interference cancellation}
\newacronym{sm}{SM}{spatial modulation}
\newacronym{snr}{SNR}{signal to noise ratio}
\newacronym{siso}{SISO}{single-input-single-output}
\newacronym{sinr}{SINR}{signal to interference and noise ratio}

\newacronym{tm}{TM}{transmission mode}

\newacronym{ul}{UL}{uplink}

\newacronym{cdf}{CDF}{cumulative  distribution  function}
\newacronym{ccdf}{CCDF}{complementary cumulative  distribution  function}
\newacronym{bw}{BW}{bandwidth}

\newacronym{iui}{IUI}{inter-user interference}
\newacronym{ue}{UE}{user equipment}
\newacronym{iai}{IaCI}{intracell interference}
\newacronym{ici}{ICI}{intercell interference}
\newacronym{re}{RE}{resource element}
\newacronym{fsic}{FSIC}{flexible successive interference cancellation}
\newacronym{pnoma}{pNOMA}{partial non-orthogonal multiple access}
\newacronym{scp}{SCP}{spatial coverage probability}
\newacronym{ccp}{CCP}{conditional coverage probability}

\newacronym{md}{MD}{meta distribution}
\newacronym{ppp}{PPP}{Poisson point process}
\newacronym{ra}{RA}{resource allocation}
\newacronym{sir}{SIR}{signal to interference ratio}
\newacronym{mgf}{MGF}{moment generation function}
\newacronym{pgfl}{PGFL}{probability generating
functional}

\newacronym{rdp}{RDP}{relative distance process}

\newacronym{tmt}{TMT}{threshold minimum throughput}

\begin{document}

\title{\linespread{1}\selectfont Meta Distribution of Partial-NOMA} % in Poisson Networks} %for Cell Center Users}

\author{
%\IEEEauthorblockN{\large  Konpal Shaukat Ali, \textit{Member, IEEE} and Arafat Al-Dweik, \textit{Senior Member, IEEE}}

\IEEEauthorblockN{\large  Konpal Shaukat Ali, \textit{Member, IEEE}, Arafat Al-Dweik, \textit{Senior Member, IEEE}, Ekram Hossain, \textit{Fellow, IEEE}, and Marwa Chafii, \textit{Member, IEEE}}

\thanks{\linespread{1}\selectfont K. S. Ali and M. Chafii are with the Engineering Division, New York University (NYU) Abu Dhabi, 129188, UAE. (Email: \{konpal.ali, marwa.chafii\}@nyu.edu). M. Chafii is also with NYU WIRELESS, NYU Tandon School of Engineering, Brooklyn, 11201, NY.\\
A. Al-Dweik is with the Center for Cyber-Physical Systems (C2PS), Department of Electrical Engineering and Computer Science, Khalifa University, Abu Dhabi, P.O.Box 127799, UAE, (Email: arafat.dweik@ku.ac.ae, dweik@fulbrightmail.org).\\
E. Hossain is with the Department of Electrical and Computer Engineering, University of Manitoba, Winnipeg, Canada (Email: ekram.hossain@umanitoba.ca).\\
This work was supported in part by the Technology Innovation Institute (TII), Abu Dhabi, UAE, Grant. no. GC/090-20.}}
\linespread{1}
\maketitle

%75- to 100-word WCL abstract limit
%75- to 200-word abstract TCOM 
%currently 100
\vspace{-15mm}
\begin{abstract} 
%\scalefont{0.99}
\textbf{This work studies the \gls{md} in a two-user \gls{pnoma} network. Compared to NOMA where users fully share a resource-element, \gls{pnoma} allows sharing only a fraction $\alpha$ of the resource-element. The \gls{md} is computed via moment-matching using the first two moments where reduced integral expressions are derived. Accurate approximates are also proposed for the $b{\rm th}$ moment for mathematical tractability. We show that in terms of percentile-performance of links, \gls{pnoma} only outperforms NOMA when $\alpha$ is small. Additionally, \gls{pnoma} improves the percentile-performance of the weak-user more than the strong-user highlighting its role in improving fairness.
  }

%This work studies the \gls{md} in a partial-non-orthogonal multiple access (pNOMA) network to obtain fine-grained information about network performance. As the \gls{md} is frequently approximated via moment matching, the first two moments of the \gls{md} are required; reduced integral expressions are derived for these. Accurate approximates are also proposed for the $n^{\rm th}$ moment to simplify the calculation. We show that in terms of percentile performance of links, pNOMA only outperforms traditional NOMA when the overlap is small. Additionally, pNOMA improves the percentile performance of weak users more than stronger users highlighting its role in improving fairness.
\end{abstract}

\glsresetall
%\begin{IEEEkeywords}

\glsresetall
\scalefont{0.98}
\section{Introduction}\label{sec1}
While \gls{noma} enables a complete sharing of the \gls{re} by multiple \gls{ue} to improve throughput, the introduced \gls{iai} deteriorates the \gls{ue} coverage. On the contrary, \gls{oma} has no \gls{iai}, which results in superior coverage, but no spectrum reuse can be applied, which hinders the throughput. Therefore, \gls{pnoma}  is introduced as a compromise between \gls{oma} and \gls{noma} \cite{myPartialNOMA}. In contrast to \gls{noma},  \glspl{ue} in \gls{pnoma} share only a fraction $\alpha$ of the \gls{re}, which allows spectrum reuse while limiting the \gls{iai} encountered by \glspl{ue}. In \cite{myPartialNOMA}, partial sharing of an \gls{re} by two users is accomplished by having the two signals overlap only with a fraction of each other in the frequency domain while having complete access to the entire time slot. The partial overlap in the frequency domain allows using matched filtering at the receiver side to further suppress the interference encountered by the \glspl{ue}, resulting in improved coverage. The matched filtering also enables devising a new decoding technique referred to as \gls{fsic}. Using matched filtering in conjunction with \gls{fsic} allows \gls{pnoma} to outperform \gls{noma} in terms of throughput \cite{myPartialNOMA}.

Stochastic geometry provides a unified mathematical paradigm for modeling large wireless networks and characterizing their operation while taking into account the experienced \gls{ici}  \cite{MH_Book2,h_tut}. Research on \gls{noma} \cite{myNOMA_tcom,my_nomaMag,N6,N18,N16,myNOMA_meta,metanoma_EH,metanoma3gpp} and \gls{pnoma} \cite{myPartialNOMA} have used these tools to analyze networks that encounter both \gls{iai} and \gls{ici}. Stochastic geometry-based studies of networks often focus on the spatial averages of performance metrics, the most frequent being the averaged \gls{scp}, which averages performance overall fading, activity, and network realizations. However, the actual performance distribution of the majority of the links may not necessarily be close to the \gls{scp}. Spatial averages thus do not reveal information about the percentile performance of links that the network operators would be interested in as these reveal the quality of service that the network can offer. It is thus pertinent to study the percentile performance of \glspl{ue}, where the fading and activity change while the network realization is kept constant. The coverage probability given a fixed network realization is defined as the \gls{ccp} \cite{meta_mh_bipolarAndCell}. The \gls{ccdf} of the \gls{ccp}, denoted as the \gls{md}, reveals the percentile performance across an arbitrary network realization. {Since the \gls{md} can only be obtained numerically using the Gil-Pelaez theorem, the beta distribution using moment matching of the first two moments of the \gls{ccp} was proposed as an accurate approximation of the \gls{md} \cite{meta_mh_bipolarAndCell}.} Works such as \cite{myNOMA_meta,metanoma_EH,metanoma3gpp}, have studied the \gls{md} for \gls{noma} \glspl{ue} {using this approach}. 

{The \gls{scp}, which is the $1{\rm st}$ moment of the \gls{ccp}, is studied in \cite{myPartialNOMA} for a two-user \gls{pnoma} network. In contrast, this work focuses on studying the \gls{md} of \glspl{ue} in such a network}  to reveal more fine-grained information about the performance of such a setup. {We derive integral expressions for the $b{\rm th}$ moment of the \gls{ccp} in a \gls{pnoma} network. For the first two moments, which are required to approximate the \gls{md} via moment matching, we are able to reduce the integrations required. We also propose accurate approximations for the $b{\rm th}$ moment of the \gls{ccp} to further simplify the integral calculation and reveal performance trends. {Using the exact and approximate moments of the \gls{ccp}, we compute the \gls{md}. This reveals fine-grained information such as the 5\%-user performance, which is the reliability that 95\% of the \glspl{ue} can achieve.} Our results highlight the superiority of {the percentile performance of} \gls{pnoma} over traditional \gls{noma} in the low $\alpha$ regime. Further, we show that \gls{pnoma} improves the performance of the weak \gls{ue} more, highlighting its role in improving fairness. }

% For the first two moments, which are required to approximate the \gls{md}, we are able to reduce the required integrals. Moreover, We propose accurate approximate moments of the \gls{md} to further simplify the integral calculation.

%The rest of the paper is organized as follows. The system model is described in Section II. The signal-to-interference ratio (SIR) analysis for the \gls{md} of the coverage probability is studied in Section III. %In Section IV, the SIR analysis for physical layer security is studied. The results are presented in Section V and the paper is concluded in Section VI.

\noindent
\textit{Notation:} We denote vectors using bold text, $\|\textbf{z}\|$ is used to denote the Euclidean norm of the vector $\textbf{z}$ and $b(\textbf{z},R)$ denotes a ball centered at $\textbf{z}$ with radius $R$. The indicator function, denoted as $\mathbbm{1}_A$ has value $1$ when event $A$ occurs and is $0$ otherwise. We use $\psi(x) =\mathrm{sin}(\pi x)/(\pi x)$ when $x \neq 0$, and $\psi(x) = 1$ when $x = 0$.
%The ordinary hypergeometric function is denoted by ${}_2F_1$,
$\mathbb{E}[\cdot]$ is the expectation and the probability of $A$ is denoted as $\mathbb{P}(A)$.

\section{System Model and Assumptions}\label{sec2}

\subsection{Network Model}\label{sec2a}
We consider a downlink cellular network where the \glspl{bs} are distributed according to a homogeneous \gls{ppp} $\Phi$ with intensity $\lambda$. {As a large network is being studied,} we assume an interference-limited regime. A \gls{bs} serves two \glspl{ue} in each \gls{re} via \gls{pnoma} using a total power budget of $P=1$. {Note that, for simplicity, we do not study \gls{pnoma} for more than two \glspl{ue} sharing a \gls{re} as such a setup complicates the analysis without providing additional insights.} {To the network, we add a \gls{bs} at the origin $\textbf{o}$, which under expectation over $\Phi$, becomes the typical \gls{bs} serving \glspl{ue} in the typical cell. We study the performance of the typical cell over one \gls{re}. As $\Phi$ does not include the \gls{bs} at $\textbf{o}$, the set of interfering \glspl{bs} for the \glspl{ue} in the typical cell is denoted by $\Phi$.} The distance between the typical \gls{bs} at $\textbf{o}$ and its nearest neighboring \gls{bs} is denoted by $\rho$. Since $\Phi$ is a \gls{ppp}, the \gls{pdf} of $\rho$ is $f_{\rho}(x)=2\pi \lambda x e^{-\pi \lambda x^2}$, $x\geq 0$.

Consider a disk around the \gls{bs} at $\textbf{o}$ with radius $\rho/2$, i.e., $b(\textbf{o},\rho/2)$, this is referred to as the in-disk \cite{myNOMA_tcom,myPartialNOMA,myNOMA_meta}. The in-disk is the largest disk centered at a \gls{bs} that fits inside its Voronoi cell. Two \gls{pnoma} \glspl{ue} are distributed uniformly and independently at random in the in-disk $b(\textbf{o},\rho/2)$ of the \gls{bs} at \textbf{o}. The rationale behind using such a model where \glspl{ue} are not too far from the serving \gls{bs} in setups where each \gls{ue} does not have an individual dedicated \gls{re} was shown in  \cite{myNOMA_meta}. A Rayleigh fading environment is assumed where the fading coefficients are \gls{iid} with a unit-mean exponential distribution. A power-law path-loss model is considered where the signal decays at the rate $r^{-\eta}$ with distance $r$, $\eta>2$ denotes the path-loss exponent and $\frac{2}{\eta}\triangleq\delta$. Fixed-rate transmissions are used by the \glspl{bs}.% where the rate of each \gls{ue} can be different.% Such transmissions result in effective rates, referred to as the throughput of the , that are lower than the transmission rate because of outage.

%\glspl{ue} outside of the in-disk are relatively far from their serving \gls{bs}, have weaker channels, and thus are better served in their own \gls{re} without sharing. Such \glspl{ue} are not considered in this letter. Therefore,

\subsection{pNOMA Model}\label{sec2b}
\begin{comment}
\begin{figure}[tb]
    \centering
    \includegraphics[width=0.4\textwidth]{TVT/figs/Blocks-1.jpg}
    \caption{Hybrid-\gls{noma} vs. \gls{pnoma}.}
    \label{fig0a}
\end{figure}
\end{comment}
\begin{comment}
\begin{figure}[tb]
%\begin{minipage}[htb]{0.7\linewidth}
\centering\includegraphics[width=0.4\textwidth]{figs/sysModel.eps}
\caption{A realization of the network. The \glspl{ue} and in-disk, marked by the dashed circle, for the cell at \textbf{o} are shown.}\label{fig0b}
%\end{minipage}\;\;\;\;
\end{figure}
\end{comment}
\begin{comment}
\begin{figure}[t]
\centering\includegraphics[width=0.4\textwidth]{figs/IalfSqrd_intrfFactorNEW.eps}
\caption{Effective interference factor $\mathcal{I}(\alpha,\beta)$ vs. $\alpha$ for different $\beta$ values using square pulses.}\label{fig0c}
%\end{minipage}
\end{figure}
\end{comment}

 A \gls{bs} serves two \glspl{ue} in each \gls{re} via \gls{pnoma} by multiplexing the signals for each \gls{ue} with different power levels using the total power $P$. %In our work two \glspl{ue} share a \gls{re}, overlapping over only a fraction of the \gls{re} and we study the performance of one such \gls{re}. Hybrid-\gls{noma} setups are not discussed further in this work.
The \gls{re} is split into three regions \cite[Fig. 1]{myPartialNOMA}, $\mathfrak{R}_1$, $\mathfrak{R}_2$ and $\mathfrak{R}_3$, where $\mathfrak{R}_2$ is shared by the two \glspl{ue}. The fraction of the \gls{bw} in $\mathfrak{R}_2$ is $\alpha$; thus, the \glspl{ue} have full access to the time slot while they share an overlap $\alpha$ of the frequency. %\footnote
%{ It should be noted that another way to achieve an overlap $\alpha$ of the \gls{re} is by having full access to the frequency channel for each \gls{ue} and only an overlap $\alpha$ of the time slot. Such an overlap scenario is not studied in this work}. 
 We refer to the fraction of the \gls{bw} in region $\mathfrak{R}_1$, accessible to only $\text{\gls{ue}}_1$, by $\beta$, where $0\leq \beta \leq 1-\alpha$. The remaining fraction of the \gls{bw}, $1-\alpha-\beta$, {in region $\mathfrak{R}_3$,} is available solely to $\text{\gls{ue}}_2$. Thus, the total fraction of the \gls{bw} available to $\text{\gls{ue}}_1$ ($\text{\gls{ue}}_2$) is $\text{BW}_1=\alpha+ \beta$ ($\text{BW}_2=1- \beta$).
 % The total fraction of the bandwidth thus available to $\text{\gls{ue}}_2$ is $\text{BW}_2=1- \beta$.% With a slight abuse of notation, in the remainder of the manuscript, we will refer to the overlap $\alpha$ in the frequency channel of the \gls{re} simply as an overlap $\alpha$ of the \gls{re}. As the entire time slot is available to both \glspl{ue}, we will disregard this aspect when referring to the partial overlap of a \gls{re}. 
 %For the special case of $\alpha=1$, \gls{pnoma} becomes traditional \gls{noma} as the \glspl{ue} overlap over the complete \gls{re}. Thus, traditional \gls{noma} is, loosely speaking, a subset of \gls{pnoma}.

An overlap in the frequency domain allows implementing filtering at the receiver side to further suppress interference. A matched filter that has a Fourier transform equal to the complex conjugate of the Fourier transform of the transmitted signal is used \cite{myPartialNOMA,AlAmmouri}. In this work, we adopt square pulse shaping for transmissions of both \glspl{ue}. A message that goes through receive filtering and has an $\alpha$-overlap with the message of the \gls{ue} of interest will be scaled by an effective interference factor $0 \leq \mathcal{I}(\alpha,\beta) \leq 1$. From \cite{myPartialNOMA}, the effective interference factor as a function of $\beta$ and the overlap $\alpha$ is calculated as 
{\footnotesize
\begin{equation}
\mathcal{I}(\alpha ,\beta )\!=\!\left(\! \int\nolimits_{\beta }^{\beta +\alpha } \!
\frac{1}{E_{1}E_{2}}\psi\!\left(\! \frac{f-f_{a}}{0.5\text{BW}_{1}}%
\!\right) \!\psi\!\left(\frac{f-f_{b}}{0.5\text{BW}_{2}} \!\right)
df \!\right) ^{{2}}
\end{equation}}
where the center frequency of $\text{\gls{ue}}_1$ is $f_a=\frac{\alpha+\beta}{2}$ and $\text{\gls{ue}}_2$ is $f_b=\frac{1+\beta}{2}$. The factors $E_i$ for $i \in \{1,2\}$ are used to scale the energy to $1$ and are calculated as $E_{i}^{2}=\int_{-\text{BW}_{i}/2}^{\text{BW}_{i}/2}\psi^{2}\left( \frac{2f}{\text{BW}_{i}}\right) df$. Note that $\mathcal{I}(\alpha,\beta)$ increases monotonically with $\alpha$ and is $1$ ($0$) when $\alpha=1$ ($\alpha=0$) \cite[Fig. 2]{myPartialNOMA}. As $0 \leq \mathcal{I}(\alpha,\beta) \leq 1$, matched filtering results in suppression of the interference from the other \gls{ue} partially sharing the \gls{re}. Since any message that has an $\alpha$-overlap with the \gls{ue} of interest is scaled by $\mathcal{I}(\alpha,\beta)$, not only does matched filtering suppress \gls{iai}, but also reduces \gls{ici}. 

%Similar to \gls{noma}, \gls{pnoma} requires ordering \glspl{ue} based on some measure of channel strength. This is required for both \gls{ra} and decoding. 
We order the \glspl{ue} based on the link distance $R$ between the typical \gls{bs} at $\textbf{o}$ and its \glspl{ue} that are uniformly distributed in the in-disk with radius $\rho$. %, the link distance is thus conditioned on $\rho$.
This is equivalent to ordering based on the decreasing received mean signal power, i.e., $R^{-\eta}$. Thus, we refer to the strong (weak) \gls{ue} as $\text{\gls{ue}}_1$ ($\text{\gls{ue}}_2$). As the order of the \glspl{ue} is known at the \gls{bs}, we use ordered statistics for the \gls{pdf} of $R_i$   \cite{N16_18},
{\footnotesize
\begin{equation}
f_{R_{i}\mid \rho }(r\mid \rho )=16r\rho ^{-2}\left( 4r^{2}\rho ^{-2}\right)
^{i-1}\left( 1-4r^{2}\rho ^{-2}\right) ^{2-i}.  \label{f_Ri}
\end{equation}}
While \gls{noma} uses \gls{sic} for decoding,
after matched filtering in \gls{pnoma}, the message of $\mathrm{\gls{ue}}_2$ scaled by $\mathcal{I}(\alpha,\beta)$ can be too weak for $\mathrm{\gls{ue}}_1$ to decode; \gls{fsic} was thus introduced in \cite{myPartialNOMA} to combat this problem.% In particular, $\text{\gls{ue}}_1$ using \gls{fsic} can decode its own message in either of two ways: 1) Similar to conventional \gls{sic}, %, the message of $\text{\gls{ue}}_2$ is first decoded, treating the message of $\text{\gls{ue}}_1$ as noise, and removed, followed by decoding of the message of $\text{\gls{ue}}_1$, or 2) the message of $\text{\gls{ue}}_1$ is decoded while treating the interference from the message of $\text{\gls{ue}}_2$ as noise. Decoding for $\text{\gls{ue}}_2$ in \gls{fsic}, as in \gls{sic}, involves simply decoding its own message while treating the message of $\text{\gls{ue}}_1$ as noise. {Again, note that, loosely speaking, \gls{sic} is a subset of \gls{fsic}.}

The power allocated to $\text{\gls{ue}}_1$ ($\text{\gls{ue}}_2$) is denoted by $P_1$ ($P_2$) and $P_1+P_2=1$. For fixed rate transmission, the rate of $\text{\gls{ue}}_i$, $i\in \{1,2\}$, is $\log(1+\theta_i)$. Accordingly, to decode the message of $\text{\gls{ue}}_i$, the \gls{sir} needs to exceed $\theta_i$.% While \gls{sic} requires the message of $\text{\gls{ue}}_2$ to be decoded by both \glspl{ue} all the time, \gls{fsic} requires the message of $\text{\gls{ue}}_2$ to be decoded by both \glspl{ue} some of the time. Thus, as in the case of \gls{sic}, \gls{fsic} allocates resources so that the message of $\text{\gls{ue}}_2$ is easier to decode by allocating it higher power and/or lower transmission rate. It should be noted that while the two \glspl{ue} only have an $\alpha$ overlap in the \gls{re}, since the power allocated to a \gls{ue} in a \gls{re} is fixed over the \gls{re} and as the sum power of the two \glspl{ue} can never exceed the power budget of $P=1$, we have $P_1+P_2=1$.%\footnote
The throughput of $\text{\gls{ue}}_i$, $i \in \{1,2\}$ is defined as $\mathcal{T}_i=\text{BW}_i \; \mathbb{P}(C_i) \; \log(1+\theta_i),$ 
where $C_i$ is the event that $\text{\gls{ue}}_i$ is in coverage. The sum throughput of the typical cell is thus $\mathcal{T}_1+\mathcal{T}_2$. As $\text{BW}_i$ is a function of both $\alpha$ and $\beta$, the resources to be allocated in a \gls{pnoma} setup for a given $\alpha$ are $P_1=(1-P_2)$, $\theta_1$, $\theta_2$ and $\beta$.  

\subsection{SIRs Associated With pNOMA and Coverage Events}\label{sec2c}

Because \gls{pnoma} uses \gls{fsic}, there are multiple \glspl{sir} of interest. For the two-user downlink \gls{pnoma} we require ${\rm SIR}_j^i$, the \gls{sir} for decoding the $j{\rm th}$ message at $\text{\gls{ue}}_i$ where $i \leq j$ and the messages of all \glspl{ue} weaker than $\text{\gls{ue}}_j$ have been removed while the messages of all \glspl{ue} stronger than $\text{\gls{ue}}_j$ are treated as noise. In particular, 
 %\scalefont{0.99}
{\footnotesize \begin{align}
&\mathrm{SIR}_{2}^{2}=\left(  h_{2}R_{2}^{-\eta}P_{2}\right)  /\left(h_{2}R_{2}^{-\eta}P_{1}\mathcal{I}(\alpha,\beta)+\tilde{I}_{2}^{\mathrm{\o}}\right)  \\
&\mathrm{SIR}_{2}^{1}=\left(  h_{1}R_{1}^{-\eta}P_{2}\mathcal{I}(\alpha,\beta)\right)  /\left(  h_{1}R_{1}^{-\eta}P_{1}+\tilde{I}_{1}^{\mathrm{\o }}\right)  \\
&\mathrm{SIR}_{1}^{1}=\left(  h_{1}R_{1}^{-\eta}P_{1}\right)  /\tilde{I}_{1}^{\mathrm{\o }}
\end{align}}
where $\tilde{I}^{\rm\o}_i$ is the \gls{ici} experienced at \text{\gls{ue}}$_i$, $\tilde{I}^{\rm\o}_i= \left(P_i +(1-P_i) \mathcal{I}(\alpha, \beta) \right)\sum_{\textbf{x} \in \Phi}  g_{\textbf{y}_i} {\|\textbf{y}_i\|}^{-\eta}$,
where $\textbf{y}_i=\textbf{x}-\textbf{u}_i$ and $\textbf{u}_i$ is the location of $\text{\gls{ue}}_i$. The fading coefficient from the serving \gls{bs} (interfering \gls{bs}) located at {$\textbf{o}$ ($\textbf{x}$) to $\text{\gls{ue}}_i$ is $h_i$ ($g_{\textbf{y}_i}$)}. The \gls{ici} scaled to unit transmission power by each interferer is defined as $I^{\rm\o}_i$, hence, $\tilde{I}^{\rm\o}_i= \left(P_i + (1-P_i) \mathcal{I}(\alpha, \beta) \right) I^{\rm\o}_i$. Because $\left(P_i+(1-P_i)\mathcal{I}(\alpha, \beta) \right)\!\!\leq \!\!1$, \gls{ici} in \gls{pnoma} is lower than its traditional counterparts. Additionally, since the network model conditions an interferer to exist at a distance $\rho$ from the typical \gls{bs} at $\textbf{o}$, we can rewrite $I^{\rm\o}_i$ as
{\footnotesize\begin{equation}
I_{i}^{\mathrm{\o }}=\sum_{\mathbf{x}\in\Phi,\Vert\mathbf{x}\Vert>\rho
}g_{\mathbf{y}_{i}}{\Vert\mathbf{y}_{i}\Vert}^{-\eta}+\sum_{\mathbf{x}\in
\Phi,\Vert\mathbf{x}\Vert=\rho}g_{\mathbf{y}_{i}}{\Vert\mathbf{y}_{i}\Vert
}^{-\eta}.\label{scaledIinterRewrite}%
\end{equation}}
{Note that as there is no interfering \gls{bs} inside $b(\textbf{o},\rho) $, the nearest interfering \gls{bs} from $\text{\gls{ue}}_i$ is at least $\rho-R_i$ away. As $\rho-R_i>R_i$, the in-disk model offers a larger guard zone than the usual guard zone of link distance for \glspl{ue} in a downlink Poisson network \cite{myNOMA_tcom}.}

 While ${\rm SIR}_1^1$ is associated with $\text{\gls{ue}}_1$ decoding its message after decoding and cancelling the message of $\text{\gls{ue}}_2$, \gls{fsic} also allows $\text{\gls{ue}}_1$ to decode its own message while treating the message of $\text{\gls{ue}}_2$ as noise. The \gls{sir} associated with $\text{\gls{ue}}_1$ for decoding its own message when the message of $\text{\gls{ue}}_2$ has not been canceled is
{\footnotesize \begin{equation}
\mathrm{\widetilde{SIR}}_{1}^{1}= \left( h_{1}R_{1}^{-\eta}P_{1} \right) /\left(  h_{1}%
R_{1}^{-\eta}P_{2}\mathcal{I}(\alpha,\beta)+\tilde{I}_{1}^{\mathrm{\o }%
}\right).\label{tildeSIR_11}%
\end{equation}}
As \gls{fsic} decoding for $\text{\gls{ue}}_2$ involves decoding its own message while treating the interference from the message of $\text{\gls{ue}}_1$ as noise, the event of successful decoding at $\text{\gls{ue}}_2$ is defined as 
{\footnotesize\begin{align}
C_2 = \left\lbrace {\rm SIR}_2^2>\theta_2 \right\rbrace = \left\lbrace h_2  >  R_2^{\eta} \tilde{I}^{\rm\o}_2   \bar{M}_2  \right\rbrace,
\label{C2}
\end{align}
\begin{equation}
\bar{M}_2= {\theta_2}/\left(\:{P_2 - \theta_2 P_1 \mathcal{I}(\alpha, \beta)} \: \right). 
\label{Mbar_2}
\end{equation}}
\gls{fsic} decoding for $\text{\gls{ue}}_1$, on the other hand, is the joint event as described in Section~\ref{sec2b}. The event of successful decoding at $\text{\gls{ue}}_1$ is thus defined as %\scalefont{0.95} 
{\footnotesize\begin{align}
&C_1= \left\lbrace \left({\rm SIR}_2^1 > \theta_2 \bigcap {\rm SIR}_1^1 > \theta_1 \right) \bigcup {\rm \widetilde{SIR}}_1^1>\theta_1 \right\rbrace \nonumber \\ &=\left\lbrace \!  h_1\!>\! R_1^{\eta} \tilde{I}^\mathrm{\o}_1   M_1   \bigcup   h_1> R_1^{\eta} \tilde{I}^\mathrm{\o}_1  M_0 \! \right\rbrace  = \left\lbrace \! h_1  >  R_1^{\eta} \tilde{I}^{\rm\o}_1  \bar{M}_1 \! \right\rbrace, \label{C1}
\end{align} 
\begin{multline}
\bar{M}_{1}=\min \left\{ M_{0},M_{1}\right\} \mathbbm{1}_{\tilde{P}_{1}>0}%
\mathbbm{1}_{\tilde{P}_{2}^{1}>0}\mathbbm{1}_{P_{1}>0}  \label{Mbar_1} \\
+M_{0}\mathbbm{1}_{\tilde{P}_{1}>0}\mathbbm{1}_{\tilde{P}_{2}^{1}\leq
0\;\cup P_{1}\leq 0}\;+\;M_{1}\mathbbm{1}_{\tilde{P}_{1}\leq 0}\mathbbm{1}_{%
\tilde{P}_{2}^{1}>0}\mathbbm{1}_{P_{1}>0}
\end{multline} }
using $\tilde{P}_1=P_1 - \theta_1 P_2 \mathcal{I}(\alpha, \beta)$, $\;\tilde{P}_2^1= P_2 \mathcal{I}(\alpha, \beta) - \theta_2 P_1 $, $\;M_0={\theta_1}{/ \tilde{P}_1}$ and $M_1= \max \left\lbrace {\theta_2}/{\tilde{P}_2^1} , {\theta_1}/{P_1} \right\rbrace$.

The event of successful decoding at $\text{\gls{ue}}_i$ is thus of the form $C_i=\left\lbrace h_i  >  R_i^{\eta} \tilde{I}^{\rm\o}_i  \bar{M}_i \right\rbrace$. Using $\tilde{I}^{\rm\o}_i=(P_i +(1-P_i) \mathcal{I}(\alpha,\beta)) I^{\rm\o}_i$ and $\tilde{M}_i=(P_i +(1-P_i) \mathcal{I}(\alpha,\beta)) \bar{M}_i $, we can rewrite $C_i$ as
{\footnotesize\begin{align}
C_i=\left\lbrace h_i  >  R_i^{\eta} I^{\rm\o}_i  \tilde{M}_i \right\rbrace. \label{Ci}
\end{align}}

\section{Analysis of the Meta Distribution}\label{sec3}
%The network operators are often interested in the percentile performance of \glspl{ue}, where the fading and activity, if any, change while the network realization is kept constant. In this regard, 
The \gls{ccp} is defined as the coverage probability given a fixed network realization \cite{meta_mh_bipolarAndCell}. For a fixed, yet arbitrary, realization of the network $\mathcal{P}_{C_i}$, the \gls{ccp} of $\text{\gls{ue}}_i$ in a \gls{pnoma} network is %defined as  $\mathcal{P}_{C_{i}}\ \overset{\Delta}{=}\mathbb{P}(C_{i}\mid\Phi)$, where
{\footnotesize\begin{align}
&  \mathcal{P}_{C_{i}} \!\overset{\Delta}{=}\mathbb{P}(C_{i}\mid\Phi) \! \overset{(a)}{=}\mathbb{E}_{g_{\mathbf{y}_{i}}} \! \left[\!
\ \exp\left( \! -R_{i}^{\eta}\tilde{M}_{i}\sum\nolimits_{\mathbf{x}\in{\Phi}%
}g_{\mathbf{y}_{i}}{\Vert\mathbf{y}_{i}\Vert}^{-\eta} \! \right)  \mid\Phi  \right]
\nonumber\\
&  \overset{(b)}{=}\prod\nolimits_{\mathbf{x}\in{\Phi}}\left(  {1+R_{i}^{\eta
}\tilde{M}_{i}{\Vert\mathbf{y}_{i}\Vert}^{-\eta}}\right)  ^{-1},\label{CCP}%
\end{align} }
where $(a)$ follows by using the definition of $C_i$ in \eqref{Ci} and the \gls{cdf} of $h_i\sim \exp(1)$. Using the \gls{mgf} of the independent random variables $g_{\textbf{y}_i} \sim \exp(1)$, $(b)$ is obtained. 

The requirement for more fine-grained information on performance leads to the notion of studying the distribution of the \gls{ccp}. The \gls{md} was thus defined as the \gls{ccdf} of the \gls{ccp} \cite{meta_mh_bipolarAndCell}. The \gls{md} for $\text{\gls{ue}}_i$ can be written as: $\bar{F}_{\mathcal{P}_{C_i}}(\mu)\stackrel{\triangle}= \mathbb{P}(\mathcal{P}_{C_i} > \mu), \;\; 0\leq \mu \leq 1. $
%\begin{align*}
%\bar{F}_{\mathcal{P}_{C_i}}(\mu)\stackrel{\triangle}= \mathbb{P}(\mathcal{P}_{C_i} > \mu), \;\;\; 0\leq \mu \leq 1. 
%\end{align*}

{The $b{\rm th}$ moment of the \gls{ccp} of $\text{\gls{ue}}_i$, by definition, can be calculated using \eqref{CCP} as
{\footnotesize\begin{equation}
\mathcal{M}_{i,b}=\mathbb{E}\left[  \mathcal{P}_{C_{i}}^{b}\right]
=\mathbb{E}\left[  \prod\nolimits_{\mathbf{x}\in{\Phi}}\left(  1+R_{i}^{\eta
}\tilde{M}_{i}{\Vert\mathbf{y}_{i}\Vert}^{-\eta}\right)  ^{-b}\right]
.\label{M_ib_gen}%
\end{equation}}
Note that by definition, the \gls{scp} of $\text{\gls{ue}}_i$ is the first moment of the \gls{ccp} of $\text{\gls{ue}}_i$, i.e., $\mathcal{M}_{i,1}$.
}

Deriving the \gls{md} is generally intractable. Therefore, the beta distribution using moment matching was proposed as a very accurate approximation for the \gls{md} \cite{meta_mh_bipolarAndCell}. This approach only requires the first two moments of the \gls{ccp}, i.e., $\mathcal{M}_{i,1}$ and $\mathcal{M}_{i,2}$. In particular,
{\footnotesize\begin{equation}
{\bar{F}_{{\mathcal{P}_{C_{i}}}}(\mu )}\approx 1-\mathcal{I}_{\mu }\left(
\beta _{i}\mathcal{M}_{i,1}\left( 1-\mathcal{M}_{i,1}\right) ^{-1},\beta
_{i}\right) 
\end{equation}}
where $\beta_i$=$\frac{(\mathcal{M}_{i,1}-\mathcal{M}_{i,2})(1-\mathcal{M}_{i,1})}{\mathcal{M}_{i,2}-\mathcal{M}_{i,1}^2}$ and $\mathcal{I}_{\mu}(a,b)=\int_0^{\alpha} l^{a-1} (1-l)^{b-1} dl$ is the regularized incomplete beta function. 
%The variance {of the \gls{md} of $\text{\gls{ue}}_i$} is defined as $\sigma_i^2={\mathcal{M}_{i,2}-\mathcal{M}_{i,1}^2}$.
%\begin{align}
%\sigma_i^2={\mathcal{M}_{i,2}-\mathcal{M}_{i,1}^2}.
%\end{align}

\subsection{Exact Moments of the Conditional Coverage Probability}\label{sec3a}
This subsection evaluates the moments of the \gls{ccp} of $\text{\gls{ue}}_i$.

\textbf{\emph{Lemma 1:}} The $b{\rm th}$ moment of the \gls{ccp} of $\text{\gls{ue}}_i$ is %\scalefont{0.97}
{\footnotesize \begin{equation}
\mathcal{M}_{i,b}\approx \mathbb{E}_{\rho ,R_{i}}\left[ \exp \left( -2\pi
\lambda \mathcal{A}_{i,b}\right) {\left( 1+\tilde{M}_{i}{R_{i}^{\eta }}{\rho
^{-\eta }}\right) }^{-b}\right],   \label{M_ib}
\end{equation}
\begin{equation}
\mathcal{A}_{i,b}=\int\nolimits_{\rho -R_{i}}^{\infty }\left( 1-\left( 1+%
\tilde{M}_{i}{R_{i}^{\eta }}{r^{-\eta }}\right) ^{-b}\right) rdr.
\end{equation}}
%\scalefont{1.02}
\textbf{\emph{Proof:}} By writing $\mathcal{M}_{i,b}$ according to the definition in \eqref{M_ib_gen}, and separating the \gls{ici} along the lines of \eqref{scaledIinterRewrite}, we obtain
{\footnotesize\begin{align*}
\mathcal{M}_{i,b} & \!= \!\mathbb{E} \! \Bigg[ \! \prod\limits_{\substack{\textbf{x}\in\Phi  \\ {\|\textbf{x} \|}>\rho}}   \!\!\left(\! 1 \!+\! {\tilde{M}_i R_i^{\eta}}{{\|\textbf{y}_i\|}^{-\eta}} \! \right)^{-b} \!\!  \prod\limits_{\substack{\textbf{x}\in\Phi  \\ {\| \textbf{x} \|}=\rho}}  \!\! \left(\! 1 \!+\! {\tilde{M}_i R_i^{\eta}}{{\|\textbf{y}_i\|}^{-\eta}} \!  \right)^{-b}      \!  \Bigg]\\
&  {\stackrel{(a)}=\mathbb{E}_{\rho, R_i} \left[ e^{-2\pi \lambda \mathcal{A}_{i,b} }  \mathbb{E}_w \left[ \! {\left(\! 1+ {\tilde{M}_i R_i^{\eta}}{{w}^{-\eta}} \! \right)}^{-b}  \! \right] \right]} \\ &{\stackrel{(b)}\approx \mathbb{E}_{\rho, R_i}  \left[ e^{-2\pi \lambda \mathcal{A}_{i,b} }    {\left(1+ \tilde{M}_i {R_i^{\eta}}{{\mathbb{E} [w]}^{-\eta}}  \right)}^{-b}   \right].}
\end{align*}}
{Using \eqref{scaledIinterRewrite}, the second term inside the expectation comes from the nearest interferer from $\textbf{o}$ which is a distance $\rho$ from $\textbf{o}$, while the first term comes from the other interferers that are located at distances larger than $\rho$ from $\textbf{o}$. In $(a)$, we arrive at the first term of this expression and \eqref{M_ib}} using the \gls{pgfl} of the \gls{ppp}. Note that since the in-disk model allows a larger guard zone, the lower limit on the distance from the nearest interferer, in the inner integral $\mathcal{A}_{i,b}$, is $\rho-R_i$. {The second term of $(a)$ comes from denoting the distance between $\text{\gls{ue}}_i$ and the \gls{bs} at distance $\rho$ from \textbf{o} by $w$. For simplicity, in $(b)$ we used the approximate mean of the distance $w$ which was validated as a tight approximation and in \eqref{M_ib} $\mathbb{E}[w] \approx \rho$ as the difference between $\mathbb{E}[w]$ and $\rho$ is less than 3.2\%  \cite{myNOMA_tcom}. }
\qed

%Since the average position of the typical $\text{\gls{ue}}_i$ is \textbf{o}, $\mathbb{E}[w] \approx \rho$ is used in \eqref{M_ib}. An exact evaluation showed that the difference between $\mathbb{E}[w]$ and $\rho$ is less than 3.2\% \cite{myNOMA_tcom,myNOMA_icc}.

For general $b$, $\mathcal{M}_{i,b}$ requires a triple integral according to \eqref{M_ib}. However, for $b \in \{1,2\}$, closed-form expressions for $\mathcal{A}_{i,b}$ in \eqref{M_ib} can be obtained, reducing the calculation in \eqref{M_ib} by one integral. Note that $\mathcal{M}_{i,1}$ and $\mathcal{M}_{i,2}$ are the two most relevant moments of the \gls{ccp} of $\text{\gls{ue}}_i$ as they are sufficient to evaluate the \gls{md} of $\text{\gls{ue}}_i$. 
%While $\mathcal{M}_{i,b}$ for general $b$ requires a triple integration according to \eqref{M_ib}, for the cases of $b=1$ and $2$ we are able to reduce the calculation by one integration. In particular, closed-form expressions for $\mathcal{A}_{i,b}$ in \eqref{M_ib} can be obtained for $b \in \{1,2\}$. It should be noted that $\mathcal{M}_{i,1}$ and $\mathcal{M}_{i,2}$ are the two most relevant moments of the \gls{ccp} of $\text{\gls{ue}}_i$ as they are sufficient to evaluate the \gls{md} of $\text{\gls{ue}}_i$. 
\\\textbf{\emph{Corollary 1:}} The inner integral for calculating the first moment of the \gls{ccp}, i.e., the \gls{scp}, of $\text{\gls{ue}}_i$, $\mathcal{M}_{i,1}$, is 
{\footnotesize\begin{align}
&\mathcal{A}_{i,1}=\frac{ \tilde{M}_i R_i^{\eta} }{\eta-2}(\rho-R_i)^{2-\eta} {}_2F_1\Big(1,1 \!- \!\delta;2 \!- \! \delta;  \frac{-\tilde{M}_i R_i^{\eta}}{(\rho-R_i)^{\eta}} \Big).
\end{align}}

\textbf{\emph{Corollary 2:}} The inner integral for calculating the second moment of the \gls{ccp} of $\text{\gls{ue}}_i$, $\mathcal{M}_{i,2}$, is 
{\footnotesize
\begin{multline}
\mathcal{A}_{i,2}=\frac{\tilde{M}_{i}R_{i}^{\eta }}{\eta }\Bigg(\frac{2(\rho
\!-\!R_{i})^{\eta }+\tilde{M}_{i}R_{i}^{\eta }}{(\rho \!-\!R_{i})^{\eta }+%
\tilde{M}_{i}R_{i}^{\eta }}(\rho \!-\!R_{i})^{2-\eta } \\
+\frac{(\eta \!-\!2)\tilde{M}_{i}R_{i}^{\eta }}{2(1\!-\!\eta )}(\rho
\!-\!R_{i})^{2-2\eta }{}_{2}F_{1}\left( 1,2\!-\!\delta ;3\!-\!\delta ;-%
\tilde{M}_{i}R_{i}^{\eta }(\rho \!-\!R_{i})^{-\eta }\right)  \\
+\frac{4(\rho \!-\!R_{i})^{2-\eta }}{(\eta \!-\!2)}{}_{2}F_{1}\left(
1,1\!-\!\delta ;2\!-\!\delta ;-\tilde{M}_{i}R_{i}^{\eta }(\rho
\!-\!R_{i})^{-\eta }\right) \Bigg).
\end{multline}
}

\subsection{Approximate Moments of the Conditional Coverage Probability}\label{sec3b}
Evaluating the $b{\rm th}$ moment of the \gls{ccp} of $\text{\gls{ue}}_i$ in Section \ref{sec3a} requires a triple integral for general $b$, and a double integral for $b=1,2$. Therefore, an alternative approach to calculate the moments that is based on the \gls{rdp} is used \cite{mh_Asymptotics}. Since \gls{pnoma}  deals with ordered link distances, the ordered \gls{rdp} is used, which for $\text{\gls{ue}}_i$ is defined as: $\mathcal{R}_{i}=\left\{\mathbf{x}\in\Phi:R_{i}\Vert\mathbf{y}_{i}\Vert
^{-1}\right\} .$
%\begin{comment}
%{\footnotesize \begin{equation}
%\mathcal{R}_{i}=\left\{\mathbf{x}\in\Phi:R_{i}\Vert\mathbf{y}_{i}\Vert
%^{-1}\right\} .
%\end{equation}}
%\end{comment}

Using the ordered \gls{rdp} for $\text{\gls{ue}}_i$, we can rewrite the moments of the \gls{ccp} of $\text{\gls{ue}}_i$ in \eqref{M_ib_gen} as
{\footnotesize\begin{equation}
\mathcal{M}_{i,b}=\mathbb{E}\left[  \prod\nolimits_{z\in{\mathcal{R}_{i}}%
}\left(  1+\tilde{M}_{i}z^{\eta}\right)  ^{-b}\right]  .\label{M_ib_rdp}%
\end{equation}}
As \eqref{M_ib_gen} is in terms of the \gls{ppp} $\Phi$, evaluating $\mathcal{M}_{i,b}$ in \eqref{M_ib} requires the probability generating functional (\gls{pgfl}) of the \gls{ppp}. Accordingly, as $\mathcal{M}_{i,b}$ in \eqref{M_ib_rdp} is in terms of the \gls{rdp} $\mathcal{R}_i$, the \gls{pgfl} of $\mathcal{R}_i$ is required to evaluate $\mathcal{M}_{i,b}$. Because $R_i$ and therefore the \gls{rdp} $\mathcal{R}_i$ are conditioned on $\rho$, the \gls{pgfl} of $\mathcal{R}_i$ is also conditioned on $\rho$.
% \begin{comment}
% \textcolor{red}{By definition of the \gls{pgfl} we have
% \begin{align*}
% &  \mathcal{G}_{\mathcal{R}_{i}\mid\rho}[f]\overset{\triangle}{=}%
% \mathbb{E}\left[  \prod\nolimits_{z\in\mathcal{R}_{i}}f(z)\right]
% =\mathbb{E}\left[  \prod\nolimits_{{\mathbf{x}\in\Phi}}f\left(  R_{i}%
% \Vert\mathbf{y}_{i}\Vert^{-1}\right)  \right]  \\
% &  \overset{(a)}{=}\mathbb{E}\left[  \prod\limits_{\mathbf{x}\in\Phi
% ,\Vert\mathbf{x}\Vert>\rho}f\left(  R_{i}\Vert\mathbf{y}_{i}\Vert^{-1}\right)
% \prod\limits_{\mathbf{x}\in\Phi,\Vert\mathbf{x}\Vert=\rho}f\left(  R_{i}%
% \Vert\mathbf{y}_{i}\Vert^{-1}\right)  \right]  .
% \end{align*}}
% {By separating the \gls{ici} along the lines of \eqref{scaledIinterRewrite} and using the \gls{pgfl} of the \gls{ppp}, the \gls{pgfl} of the ordered $\mathcal{R}_i$, conditioned on $\rho$ is obtained as
% \begin{align}
% \!\mathcal{G}_{\mathcal{R}_i \mid \rho}[f] \! =\! \mathbb{E}_{R_i} \! \Bigg[\!\exp \! \Big(\!\!- \! 2 \pi \lambda \!\! \int\limits_{\rho-R_i}^\infty \! \!\! \left( \! 1 \! - \! f \! \left( \! \frac{R_i}{a} \! \right) \!\! \right) \!  a\!da \! \Big) \!\! \prod\limits_{\substack{\textbf{x}\in\Phi  \\ \| \textbf{x} \|=\rho}} \!\! f \! \left( \! \frac{R_i}{\|\textbf{y}_i\|} \! \right) \! \Bigg].\label{pgfl_rdp_CC}
% \end{align}}
% \end{comment}

\textbf{\emph{Lemma 2:}} The \gls{pgfl} of the ordered $\mathcal{R}_i$ conditioned on $\rho$ is
{\footnotesize\begin{align}
\!\mathcal{G}_{\mathcal{R}_i \!\mid  \rho}\![f] \! =\! \mathbb{E}_{R_i \!\mid  \rho} \!\Bigg[ \!\exp \! \Big(\!\!- \! 2 \pi \! \lambda \!\! \! \int\limits_{\rho \!-\!R_i}^\infty \! \!\! \left( \! 1 \! \! - \! \! f \! \left( \! \frac{\!R_i}{\!a} \! \right) \!\! \right) \!  a\!da \! \Big) \!\! \! \prod\limits_{\substack{\textbf{x}\in\Phi  \\ \| \!\textbf{x} \|=\!\rho}} \!\! f \! \left( \!\! \frac{R_i}{\| \textbf{y}_i\|} \!\! \right) \! \Bigg].  \label{pgfl_rdp_CC}
\end{align}}
\textbf{\emph{Proof:}} By definition of the PGFL we have
{\footnotesize \begin{align*}
&\mathcal{G}_{\mathcal{R}_i \mid \rho}[f]  \stackrel{\triangle}= \mathbb{E} \Big[\prod\limits_{z \in \mathcal{R}_i} f(z) \Big] = \mathbb{E} \Big[\prod\limits_{{\textbf{x} \in \Phi }} f\left({R_i}{\|\textbf{y}_i\|^{-1}}\right) \Big]. % \\
%&\stackrel{(a)}= \mathbb{E} \Bigg[\prod\limits_{\substack{\textbf{x}\in\Phi  \\ \| \textbf{x} \|>\rho}}f\left(\frac{R_i}{\|\textbf{y}_i\|}\right) \prod\limits_{\substack{\textbf{x}\in\Phi  \\ \| \textbf{x}  \|=\rho}} f\left(\frac{R_i}{\|\textbf{y}_i\|}\right) \Bigg].
\end{align*} }
By separating the interferers along the lines of \eqref{scaledIinterRewrite} and using the \gls{pgfl} of the \gls{ppp}, \eqref{pgfl_rdp_CC} is obtained. \qed  
%, (a) is obtained  into two types, the interferers that are farther than $\rho$ from \textbf{o} and the interferer conditioned to be a distance $\rho$ away from \textbf{o}. From this, we obtain the first term in \eqref{pgfl_rdp_CC} using the PGFL of the PPP. 

%However, it is infeasible to derive a  closed-form expression for \eqref{pgfl_rdp_CC}.
{However, it is infeasible to obtain tractable expressions for the moments of the CCP using \eqref{pgfl_rdp_CC}.}Therefore, we propose the use of two approximations to relax the constraints and simplify the calculation of the \gls{pgfl} of the ordered \gls{rdp} for $\text{\gls{ue}}_i$. The approximations are: 
\begin{itemize}
\item \textbf{A1}: The guard zone around the \glspl{ue} is approximated to be of radius $R_i$ (instead of the largest guaranteed radius $\rho-R_i > R_i$). %although the largest guaranteed guard zone has radius $\rho-R_i$.% As $\rho-R_i > R_i$.% this approximation overestimates the \gls{ici} encountered by the \glspl{ue}.
\item \textbf{A2}: Deconditioning on the \gls{bs} located a distance $\rho$ from $\textbf{o}$.% This approximation underestimates the \gls{ici} experienced by the \glspl{ue}.
\end{itemize}
Note that the two approximations have opposing effects on how much \gls{ici} is accounted; \textbf{A1} overestimates it while \textbf{A2} underestimates it.
%Using \textbf{A1} and \textbf{A2} we can calculate the \gls{pgfl} of $\mathcal{R}_i$ in closed-form.

\textbf{\emph{Lemma 3:}} The PGFLs of the ordered \gls{rdp} for $\text{\gls{ue}}_2$ and $\text{\gls{ue}}_1$, respectively, using approximations \textbf{A1} and \textbf{A2} are
{\footnotesize\begin{align}
&\widetilde{\mathcal{G}}_{\mathcal{R}_2 | \rho}[f] \!=\! \frac{32}{\rho^4}  \frac{\Gamma(2) \!-\! \Gamma \Big(2,\frac{\rho^2}{2}  \pi \lambda \int_1^{\infty} \left(\! 1\!-\! f(\frac{1}{y}) \!\right) y dy  \Big)} {\Big(\! 2\pi \lambda \int_1^{\infty} \left(\! 1\!-\! f(\frac{1}{y}) \!\right) y dy \! \Big)^{2}}  \label{approxG_R2} \\
&\widetilde{\mathcal{G}}_{\mathcal{R}_1 | \rho}[f] \!=\!    \frac{ \Gamma(1) \! - \! \Gamma \! \left(\! 1,\frac{\rho^2 }{2} \! \pi \lambda  \int_1^{\infty} \! \left(\! 1 \!-\! f(\frac{1}{y}) \!\right) y dy  \!\right)\! }{ \frac{\rho^2}{4}\pi \lambda \int_1^{\infty} \left(\! 1\!-\! f(\frac{1}{y}) \!\right) y dy}  \!-\!  \widetilde{\mathcal{G}}_{\mathcal{R}_2 | \rho}[f]. \label{approxG_R1}
\end{align}}
\textbf{\emph{Proof:}} Along the lines of the proof of Lemma 2, the \gls{pgfl} of the ordered \gls{rdp} for $\text{\gls{ue}}_i$ using \textbf{A1} and \textbf{A2} is
{\footnotesize\begin{align*}
&\widetilde{\mathcal{G}}_{\mathcal{R}_i \mid \rho}[f]  =\mathbb{E}_{R_i \mid \rho} \left[\exp\left(-2 \pi \lambda \int_{R_i}^\infty \left(1-f\left({R_i}/{a} \right) \right) a \; da \right)  \right].
\end{align*}}
Here the second term in \eqref{pgfl_rdp_CC} has been removed due to \textbf{A2}. Additionally, the lower limit of the integral of the first term in \eqref{pgfl_rdp_CC} is updated to reflect $R_i$, the radius of the guard zone in \textbf{A1}. Using $f_{R_i \mid \rho}$ in \eqref{f_Ri} for $i=2$ and $i=1$ we obtain \eqref{approxG_R2} and \eqref{approxG_R1}, respectively.
\qed

\textbf{\emph{Lemma 4:}} The $b{\rm th}$ moments of the \gls{ccp} for $\text{\gls{ue}}_2$ and $\text{\gls{ue}}_1$, respectively, using approximations \textbf{A1} and \textbf{A2} are
{\footnotesize
\begin{align}
&\widetilde{\mathcal{M}}_{2,b}\!=\! \mathbb{E}_{\rho} \!  \left[ \! \frac{32}{\rho^4}    \frac{ 1\!-\! \Gamma \left(2, \frac{\rho^2 }{4}  \pi \lambda \! \left(\! {}_2F_1(b,\!-\!\delta;1 \!-\! \delta; \!-  \tilde{M}_2) \!-\!1 \! \right) \right)  }{\left( \pi \lambda \! \left(\! {}_2F_1(b,-\delta;1 \!-\! \delta; -  \tilde{M}_2) -1 \!\right) \right)^{2}} \!  \right] \label{approxM2b} \\
&\widetilde{\mathcal{M}}_{1,b} = \mathbb{E}_{\rho} \Bigg[ \frac{8}{\rho^2}   \frac{1- \exp \left(- \frac{\rho^2}{4} \pi \lambda \left( {}_2F_1(b,-\delta;1 \!-\! \delta; -  \tilde{M}_1) -1\right) \right)}{\pi \lambda \left( {}_2F_1(b,-\delta;1 \!-\! \delta; -  \tilde{M}_1) -1\right) }  \nonumber \\
& - \frac{32}{\rho^4}   \frac{ 1- \Gamma \left(2, \frac{\rho^2}{4}  \pi \lambda \left( {}_2F_1(b,-\delta;1 \!-\! \delta; -  \tilde{M}_1) -1 \right)  \right) }{\left( \pi \lambda \left( {}_2F_1(b,-\delta;1 \!-\! \delta; -  \tilde{M}_1) -1 \right) \right)^{2}} \Bigg]. \label{approxM1b}
\end{align}}
\textbf{\emph{Proof:}} Using the definition in \eqref{M_ib_rdp}, we have
{\footnotesize \begin{align*}
\widetilde{\mathcal{M}}_{i,b} = \mathbb{E}_{\rho} \left[ \widetilde{\mathcal{G}}_{\mathcal{R}_i \mid \rho}[f(z)] \Big|_{f(z)=(1+\tilde{M}_i z^{\eta})^{-b}} \right].
\end{align*}}
$\widetilde{\mathcal{G}}_{\mathcal{R}_i \mid \rho}[f(z)]$ in \eqref{approxG_R1} and \eqref{approxG_R2} for $\text{\gls{ue}}_1$ and $\text{\gls{ue}}_2$, respectively, are then plugged into the equation above. We arrive at \eqref{approxM2b} and \eqref{approxM1b} using the following, where (a) is obtained using $z\rightarrow g^{-1}$:
{\footnotesize\begin{align*}
&\int_1^{\infty} \! \left(\! 1\!-\! f(z^{-1}) \!\right) z dz \Big|_{f(z)=(1+\tilde{M}_i z^{\eta})^{-b}} \\
&\!=\! \int_1^{\infty} \!\! \left(\!1 \!-\! (1 \!+\!\tilde{M}_i z^{-\eta})^{-b} \! \right) z dz \!\stackrel{(a)}=\! \frac{{}_2F_1\left(b, \!-\delta;1\!-\!\delta,\!-\tilde{M}_i \right) \!-\!1}{2}. \qed 
\end{align*}}

\section{Results}\label{sec5}
We consider \gls{bs} intensity $\lambda=10$ and $\eta=4$. Simulations are repeated {$10^5$} times. Fixed \gls{ra} is used in some of the figures while the other figures use the optimum \gls{ra} associated with a problem that aims to maximize cell sum throughput while constrained to a \gls{tmt} according to \cite[Algorithm 1]{myPartialNOMA}. Note that solving such a problem results in \gls{ra} such that the minimum required resources are spent on $\text{\gls{ue}}_2$ to attain throughput equal to the \gls{tmt} and the remaining resources are given to $\text{\gls{ue}}_1$ to maximize its throughput. The exact moments of the \gls{ccp} are used unless specified otherwise. %As $\alpha$ increases, a switch

%\subsection{Meta Distribution of Partial-\gls{noma}}\label{ResultsA}

\begin{comment} %keke
\begin{figure}[tb]
\centering
\includegraphics[width=0.4\textwidth]{figs/Kfigs/K2_MD_mu_differentP1_alpha1.eps}
\caption{\gls{md} vs. $\mu$ using $\theta_1=1$ dB, $\theta_2=0.5$ dB, and $P_1=1/3$.}
\label{fig1}
%\vspace{-5mm}
\end{figure}
\end{comment}

\begin{figure*}[htb]
\begin{minipage}[htb]{0.32\linewidth}
\centering \includegraphics[width=\textwidth]{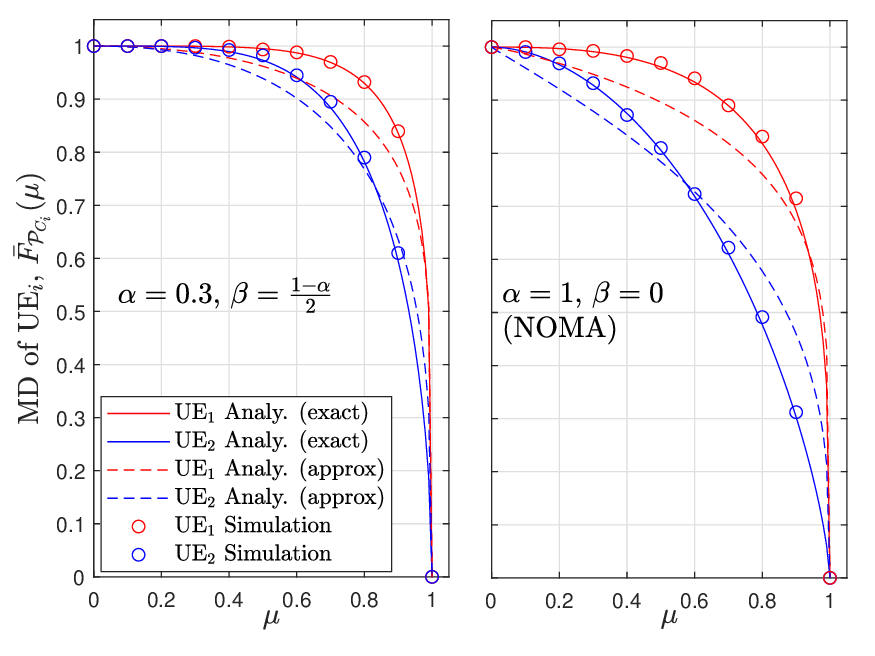}
\caption{\gls{md} vs. $\mu$ using $\theta_1=1$ dB, $\theta_2=0.5$ dB, and $P_1=1/3$.}
\label{fig1}
\end{minipage}\;\;
\begin{minipage}[htb]{0.32\linewidth}
\centering \includegraphics[width=\textwidth]{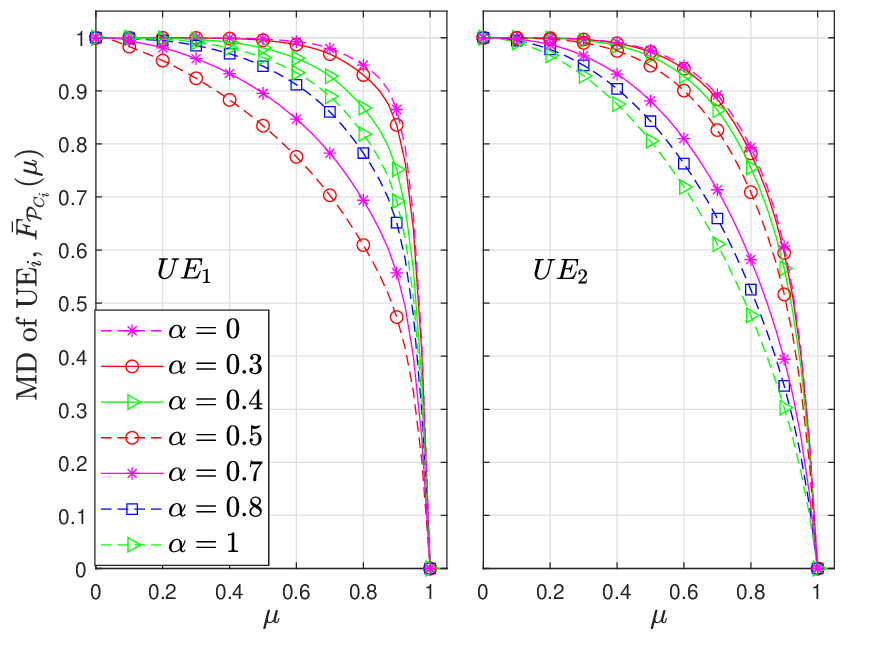}
\caption{\gls{md} vs. $\mu$, $P_1\!\!=\!\!1/3$, $\theta_1\!=\!1$ dB, $\theta_2=0.5$ dB and $\beta=\frac{1-\alpha}{2}$.}
\label{fig3}
\end{minipage}\;\;
\begin{minipage}[htb]{0.32\linewidth}
\centering \includegraphics[width=\textwidth]{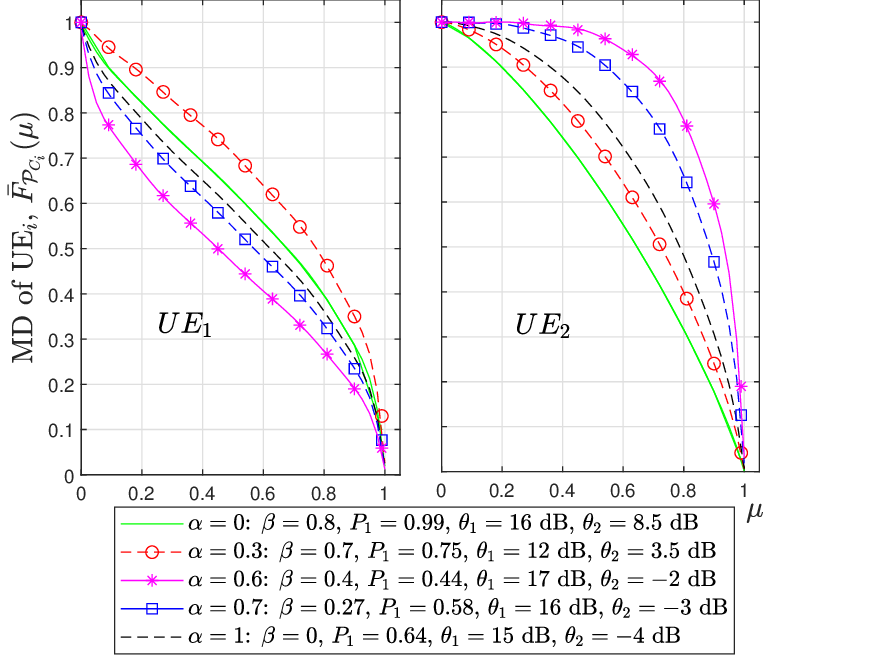}%\vspace{2mm}
\caption{MD vs. $\mu$ using optimum \gls{ra} for each $\alpha$ with \gls{tmt}$=0.25$. }
\label{fig4}
\end{minipage}
%\vspace{-3mm}
\end{figure*}

Fig. \ref{fig1} is a plot of the \gls{md}, using both the exact moments (Lemma 1) and the approximate moments (Lemma 4), for $\alpha=0.3$ and $\alpha=1$, traditional \gls{noma}, using fixed \gls{ra}. The figure validates the analysis in Section \ref{sec3} as the simulation results are a tight match with the exact moments. The approximate moments are not as close to the simulation but follow the same trends {which still gives insights on network performance at a lower computational cost}. As anticipated, due to the opposing natures of \textbf{A1} and \textbf{A2}, the approximate moments can both underestimate and overestimate the MD. Focusing on the exact moments, we observe that for $\alpha=0.3$, $93\%$ of $\text{\gls{ue}}_1$ and $78.2\%$ of $\text{\gls{ue}}_2$ achieve a reliability of $0.8$. With $\alpha=1$, on the other hand, only $81.9\%$ of $\text{\gls{ue}}_1$ and $47.7\%$ of $\text{\gls{ue}}_2$ achieve a reliability of $0.8$. Thus, a notable improvement in the percentage of \glspl{ue} able to achieve the same reliability is observed for both \glspl{ue} (13.5 \% increase for $\text{\gls{ue}}_1$ and 63.9 \% for $\text{\gls{ue}}_2$) in \gls{pnoma} compared to \gls{noma}. {We also observe that the 5\%-user performance that network operators are often interested in, defined as the reliability that 95\% of the \glspl{ue} achieve but 5\% do not, has similar trends. With $\alpha=0.3$, 95\% of $\text{\gls{ue}}_1$ ($\text{\gls{ue}}_2$) achieve a reliability of 0.76 (0.58), while in NOMA, they achieve a reliability of 0.55 (0.245).} These results emphasize the significance of the partial overlap on percentile performance.

Fig. \ref{fig3} plots the \gls{md} for different $\alpha$ values using fixed \gls{ra}. We observe that in general, when \gls{ra} is fixed, increasing $\alpha$ deteriorates the performance of $\text{\gls{ue}}_2$ due to the increasing $\mathcal{I}(\alpha,\beta)$ and consequently, interference. Thus the percentage of $\text{\gls{ue}}_2$ that can achieve at least a certain coverage probability $\mu$ decreases with $\alpha$. The performance of $\text{\gls{ue}}_1$, on the other hand, is more complex as it first deteriorates as $\alpha$ increases from $0$ to $0.5$ and then improves from $\alpha=0.5$ to $1$. {When $\alpha\leq 0.5$, $\text{\gls{ue}}_1$ treats the message of $\text{\gls{ue}}_2$ as noise; increasing $\alpha$ increases $\mathcal{I}(\alpha,\beta)$ and therefore the \gls{iai} from $\text{\gls{ue}}_2$, thereby deteriorating the performance of $\text{\gls{ue}}_1$. When $\alpha>0.5$, there is a switch in decoding technique and the message of $\text{\gls{ue}}_2$ is decoded by $\text{\gls{ue}}_1$ before decoding its own message; increasing $\alpha$ and therefore $\mathcal{I}(\alpha,\beta)$ in this case increases the power of $\text{\gls{ue}}_2$'s message making it easier to decode, thereby improving the performance of $\text{\gls{ue}}_1$. Thus, in Fig. \ref{fig3}, for $\text{\gls{ue}}_1$ \gls{pnoma} outperforms \gls{noma} ($\alpha=1$) only when $\alpha \leq 0.4$, while for $\text{\gls{ue}}_2$ \gls{pnoma} always outperforms \gls{noma}.}

\begin{comment}%keke
\begin{figure}[tb]
%\begin{minipage}[htb]{0.8\linewidth}
\centering
\includegraphics[width=0.4\textwidth]{figs/Kfigs/K_MDvsAlpha_differentAlphaBW_optRA.eps}\vspace{3mm}
\caption{MD vs. $\mu$ using optimum \gls{ra} for each $\alpha$ with \gls{tmt}$=0.25$. }
\label{fig4}
%\vspace{-3mm}
\end{figure}
\end{comment}

Fig. \ref{fig4} plots the \gls{md} for different $\alpha$ values using the optimum \gls{ra} associated with $\text{\gls{tmt}}=0.25$. Unlike Fig. \ref{fig3}, the \gls{ra} here is not the same for all $\alpha$ values {and is specified in the legend for each $\alpha$}. The performance of $\text{\gls{ue}}_2$ increases with $\alpha$ from $0$ to $0.6$. This occurs because increasing $\alpha$ increases $\text{BW}_2$, lowering the $\theta_2$ required to achieve $\text{\gls{tmt}}=0.25$. Increasing $\alpha$ beyond $0.6$ decreases the performance of $\text{\gls{ue}}_2$ as the dominance of the growing impact of $\mathcal{I}(\alpha,\beta)$ increases interference. %This happens due to the dominance of the growing impact of $\mathcal{I}(\alpha,\beta)$ increasing interference and deteriorating performance.
While $\text{\gls{ue}}_2$ aims to use the minimum resources to achieve \gls{tmt}, the goal of $\text{\gls{ue}}_1$ is to achieve the largest possible throughput. The performance in terms of the \gls{md} for $\text{\gls{ue}}_1$ initially increases from $\alpha=0$ to $0.3$, as the increasing $\text{BW}_1$ reduces the $\theta_1$ required to achieve maximum throughput. The performance decreases as $\alpha$ is increased to 0.6. {This is attributed to the switch in decoding decoding at higher $\alpha$. As $\mathcal{I}(\alpha,\beta)$ is not very high in this regime, decoding $\text{\gls{ue}}_2$'s message is difficult for $\text{\gls{ue}}_1$, and thus $\text{\gls{ue}}_1$'s performance suffers.} Increasing $\alpha$ beyond 0.6 improves performance as the higher $\mathcal{I}(\alpha,\beta)$ makes decoding $\text{\gls{ue}}_2$'s message easier for $\text{\gls{ue}}_1$ and as the increasing \gls{bw} for both \glspl{ue} allows more power to be left for $\text{\gls{ue}}_1$'s message. 

{We observe, from Figs. \ref{fig1} and \ref{fig3} explicitly, that $\text{\gls{ue}}_2$ with the appropriate $\alpha$ outperforms its NOMA counterpart more than $\text{\gls{ue}}_1$.} {This can be attributed to the additional flexibility that comes with the ability to select $\alpha$ and $\beta$ in \gls{pnoma}, which are fixed in NOMA. The observation} sheds light on the fact that with appropriate $\alpha$, \gls{pnoma} is able to assist the weaker \gls{ue} significantly more than \gls{noma} highlighting the role of \gls{pnoma} in improving \gls{ue} fairness. %\textcolor{blue}{Note that this can be attributed to the additional flexibility that comes with the ability to select $\alpha$ and $\beta$ in \gls{pnoma} which are fixed in NOMA.}

\section{Conclusion}\label{sec6}
The \gls{md} of a \gls{pnoma} network was studied to obtain fine-grained information on network performance. Integral expressions were obtained for the moments of the \gls{ccp}. We reduced the integrals for the first two moments, which are required for approximating the \gls{md} via moment matching. By proposing the use of two approximations, accurate approximate moments of the \gls{ccp} were derived that further simplified the integral calculation. We showed that in terms of percentile performance of links, \gls{pnoma} outperforms \gls{noma} only when $\alpha$ is lower than a certain value. This highlights that deploying \gls{pnoma} over \gls{noma} is only efficient when $\alpha$ is low and sheds light on careful parameter selection. Our results also indicated that \gls{pnoma} helps improve the performance of weaker \glspl{ue} highlighting its significance as a means to improve \gls{ue} fairness. 

%The impact of different parameters on the percentile performance of each \gls{ue} in a \gls{pnoma} network with different $\alpha$ was shown. .....We showed that at low $\alpha$ \gls{pnoma} outperforms \gls{noma} in terms of percentile performance of links while \gls{noma} is superior to higher $\alpha$, shedding light on parameter selection.

\linespread{1}
\bibliographystyle{IEEEtran}
\bibliography{References}

%%%%%%%%%%%%%%%%%%%%%%%%%%%%%%%

%\input{Response_Letter}
\end{document}